# Understanding visual processing of motion: Completing the picture using experimentally driven computational models of MT

**Parvin Zarei Eskikand[1*], David B Grayden[1], Tatiana Kameneva[1,2], Anthony N Burkitt[1], Michael R Ibbotson[3]**

[1] Department of Biomedical Engineering, The University of Melbourne, Parkville, Australia, 3052
[2] Faculty of Science, Engineering and Technology, Swinburne University of Technology, Hawthorn, Australia, 3122
[3] National Vision Research Institute, Australian College of Optometry, Carlton, Australia, 3053
*Corresponding author
Email: pzarei@unimelb.edu.au

# Abstract

Computational modeling helps neuroscientists to integrate and explain experimental data obtained through neurophysiological and anatomical studies, thus providing a mechanism by which we can better understand and predict the principles of neural computation. Computational modeling of the neuronal pathways of the visual cortex has been successful in developing theories of biological motion processing. This review describes a range of computational models that have been inspired by neurophysiological experiments. Theories of local motion integration and pattern motion processing are presented, together with suggested neurophysiological experiments designed to test those hypotheses.

# 1. Introduction

Motion processing in the primate brain involves many specialized areas in the visual pathways, with very complex interconnectivity between those areas (Van Essen and Maunsell 1983). The sheer complexity of these motion-specialist areas makes it very challenging to understand biological motion processing without building models that attempt to incorporate all the known elements of the biological system.

Computational modelling is, therefore, an essential tool that allows us to study the specific roles of the neurons and neural circuits in biological visual pathways and their contribution to the ultimate goal of transferring information about the incoming visual image into perceptions and actions. Computational modelling assists in developing theories of the detailed circuitry of the connections between neurons based on existing neurophysiological data and, in turn, allows us to pose hypotheses that can be investigated through experimental research. This review outlines the state of our knowledge regarding motion processing in the primate brain, the models that have allowed us to understand that processing and what comes next.

## 1.1. Motion processing in the primate brain

The first location in the primate visual pathway where many neurons with direction-selective properties are found is the primary visual cortex (V1) (Gur and Snodderly 2007), which receives most of its sensory input from the dLGN. The populations of V1 neurons can be divided into two overarching groups: simple and complex cells (Hubel and Wiesel 1965). Simple cells have RFs characterized by readily identifiable, neighboring bright-sensitive (ON) and dark-sensitive (OFF) bands. The spatial summation between ON and OFF sensitive regions within the RFs of simple V1 neurons is approximately linear. Movshon et al. (1978) suggested that the RFs of complex V1 neurons are composed of subunits whose function is very similar to simple V1 neurons. The spatial integration of the RFs of complex V1 neurons is nonlinear. Therefore, for complex cells, so long as an edge has the appropriate orientation, it doesn't matter exactly where within the complex cell's RF the ON and OFF regions that make up the edge are located.

Neurons in V1 send their information to higher centers of the cortex through two different pathways: the dorsal pathway and the ventral pathway (Goodale and Milner 1992). The dorsal pathway projects into the parietal lobe and the ventral pathway to the inferior temporal lobe (Goodale and Milner 1992). The dorsal pathway is mainly involved in determining spatial location, detecting image motion and determining the direction of motion (Fukushima and Kikuchi 1995). The ventral pathway is specialized for processing form information, including color, shape and texture (Fukushima and Kikuchi 1995).

Ever since Zeki's (1974) discovery that neurons in area V5 of the primate visual cortex (now more often referred to as the middle temporal area or MT) were specialized for detecting the direction of image motion, a great deal of effort has gone into understanding the mechanisms underlying motion processing in this primate brain region (Zeki 1974). The most numerous and, arguably, the most important direct input to area MT is from V1 (Maunsell and van Essen 1983). The connectivity map of the information flow in a subset of areas in the visual cortex is shown in Figure 1.

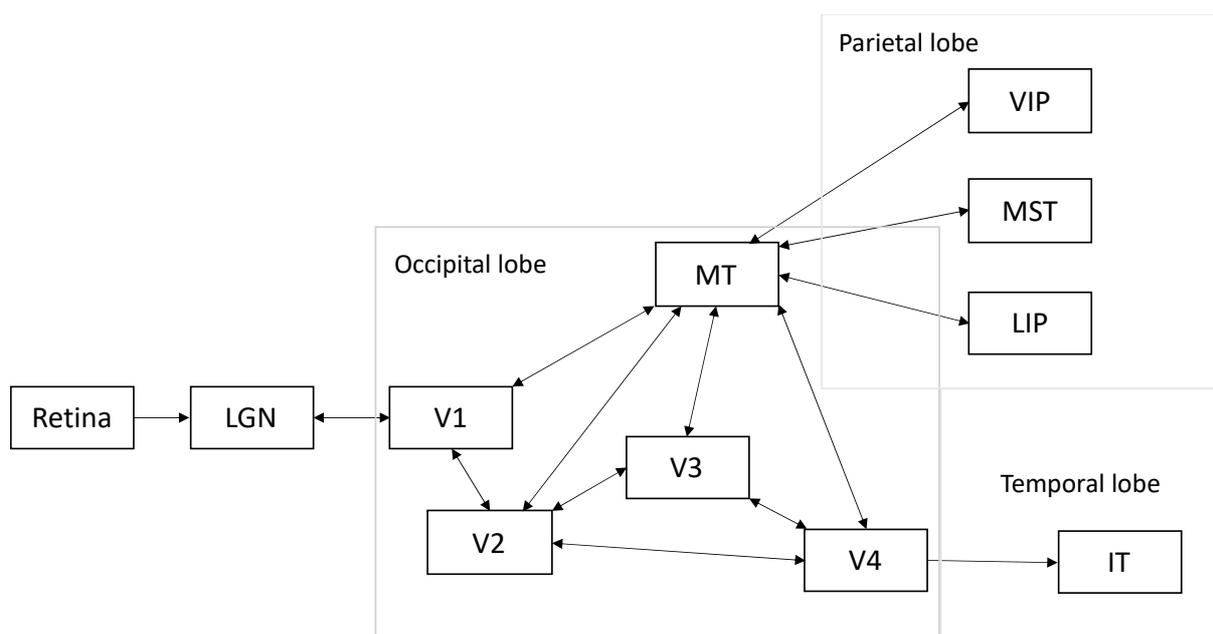

Figure 1. Connectivity map between a subset of areas in the visual cortex. The visual information received from the retina is transmitted to the primary visual cortex (V1) after passing through the Lateral Geniculate Nucleus (LGN). The V1 area receives feedback connections from areas including V2 and the Middle Temporal (MT) area. There are bidirectional connections between MT and other areas in the visual cortex, including V1, V2, V3, V4, Medial Superior Temporal (MST), Lateral Intraparietal (LIP) and Ventral Intraparietal (VIP) areas. The visual information from V4 projects to the Inferior Temporal (IT) area.

Cells in area MT have substantially larger receptive fields than those in V1 and summate the responses of multiple V1 cells (Zeki 1971). Many studies that assess the mechanisms underlying MT responses use plaid patterns. These are constructed from two drifting grating patterns (Fig. 2a, left and middle insets). The component gratings are highly oriented structures and move in different directions from each other. The direction of the pattern is formed by summation of two drifting gratings moving in different directions (Fig. 2a, right inset). When plaid patterns are presented to MT neurons, the cells show a range of response characteristics. At one end of the spectrum are cells that are tuned to detect the directions of motion of the component gratings (Fig. 2b, left). At the opposite end of the spectrum are cells that respond optimally to the direction of motion of the composite pattern, which matches human perception (Fig. 2b, right). Not surprisingly, cells at either end of the spectrum are known respectively as component and pattern cells. Importantly, around a third of MT neurons have elements of both types, showing very broad directional tuning functions that respond to both the component and pattern directions (Movshon and Newsome 1996).

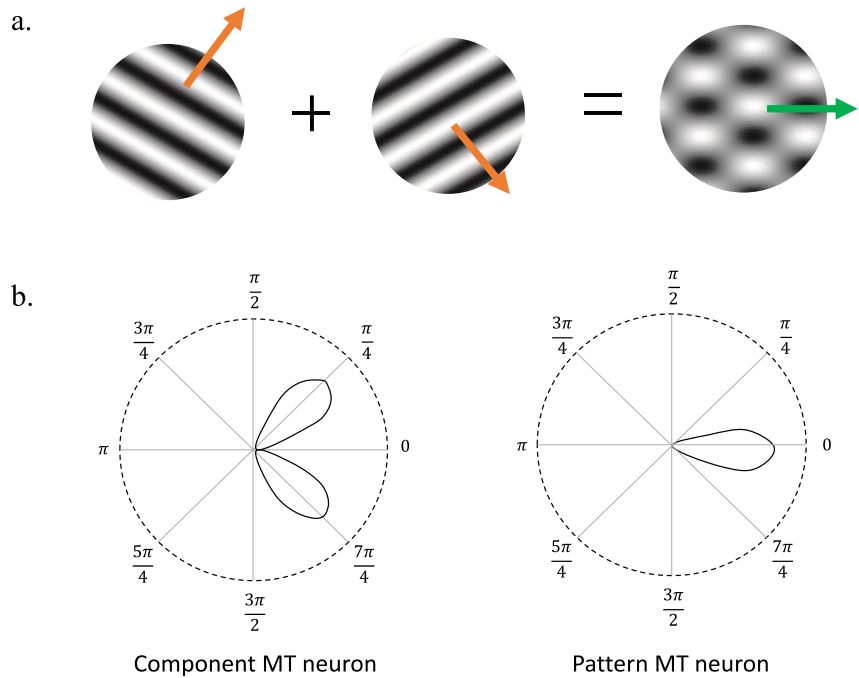

Figure 2. (a) A plaid motion made of two gratings moving in different directions. (b) A schematic of the directional tuning of component and pattern MT neurons to the plaid motion. In this type of experiment, the relative difference in motion direction between the two components is constant but the combined direction of the pattern motion is presented at all directions.

### 1.2 . Motion processing with artificial stimuli

Motion perception sounds like a simple concept but can be a surprisingly difficult task to process. A wide range of artificial stimuli have been used through the history of vision science to understand how motion information is processed in the cortex. To highlight the complexity and contextual dependence of visual motion processing, we will describe six commonly used artificial motion stimuli that exemplify different motion cues and how they are integrated contextually. These stimuli have been generally used to target specific aims in studying motion processing in the visual system. Addressing the specific aims by using these six commonly used stimuli has provided a common basis for designing neurophysiological experiments and developing computational models to better understand motion processing in the visual cortex. The neurophysiological findings in response to these stimuli have driven the development of computational models to replicate the responses of the neurons and, in turn, the computational models have inspired new hypotheses on the neural circuitry in response to these stimuli.

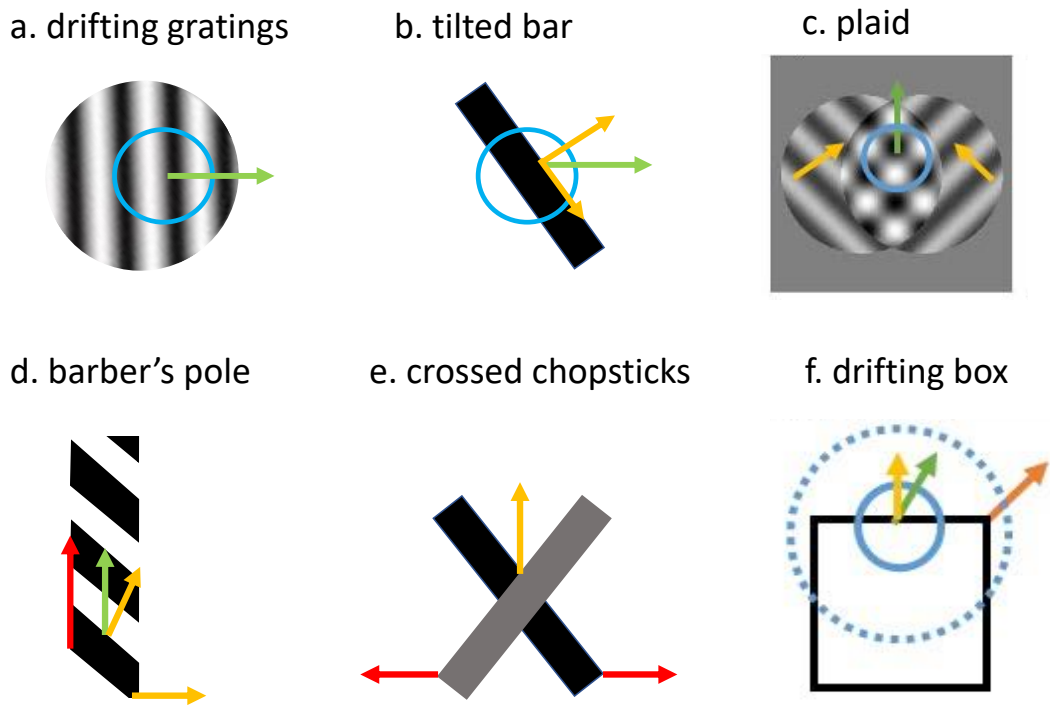

Figure 3. Six commonly used types of motion stimuli. (a) Drifting gratings. The blue circle represents the RF of the neurons and green arrow shows the direction of motion of gratings. (b) Tilted bar stimulus. The blue circle represents the RF of a neuron. Yellow arrows are the local component of the motion, perpendicular to the edge and in parallel to the edge of the bar, green arrow is the actual direction of motion of the bar. (c) Plaid stimulus. The green arrow shows the pattern motion direction and yellow arrows show the direction of motion of gratings. The blue circle represents the RF of the neuron. (d) Barber's pole stimulus. The green arrow shows the perceived global direction of motion. The red arrow shows the motion of the terminators formed at the edge of the stimulus. Yellow oblique (up and right) arrows show the local motion created perpendicular to the edge of the bar. Yellow right arrow shows the actual direction of motion. (e) Crossed chopsticks. Two crossing bars moving in the opposite directions. The motion of the extrinsic terminators (terminators formed because of overlapping with another object) is shown by the yellow arrow. The motion of intrinsic terminators is shown by the red arrow that matches with the actual direction of motion of the bars. (f) Drifting boxes. The box is moving in the up-right direction. The solid blue circle represents the center of the RF and the dotted blue circle

shows the surround. The motion at the edge of the bar is ambiguous because of the aperture problem. The local motion signal at the edge of the box is moving up (yellow arrow) direction despite the fact that the box is moving in the up-right direction (green arrow). The surround of MT neurons has facilitatory effect in this case. The motion of the terminators (orange arrow) matches the correct direction of motion. The surround of MT neurons has antagonistic effect at the terminators.

**Aperture problem rising from the response to drifting gratings or tilted bars**

Neurons in V1 in the primate brain show a pronounced tuning for oriented features such as edges. Many V1 neurons are also direction-selective, responding more to one direction of motion (perpendicular to their preferred orientation) than the opposite direction (Figure 3a, green arrow) (De Valois et al. 1982). V1 neurons can only measure local motion signals, i.e., the component of the motion that moves orthogonal to the orientation of an edge (Figure 3b, up-right yellow arrow). Any component of the motion parallel to the edge is not detectable by complex neurons because of the invariance of the contrast in this direction (Figure 3b, down-right yellow arrow). Measuring only one component of the motion results in an ambiguity of the direction of motion, called the "aperture problem". When viewed through a narrow V1 receptive field, the edge of a moving bar only appears to move perpendicular to the edge of the bar (Pack et al. 2003, Born and Bradley 2005, Bradley and Goyal 2008, Smith et al. 2009). This is illustrated in Figure 3b, where the bar is moving to the right (green arrow) but, when only viewed through the aperture (blue circle representing the RF), the portion of the bar appears to be moving perpendicular to its leading edge (up-right yellow arrow). This limitation is more evident when using drifting gratings (Figure 3a) or tilted bars (Figure 3b) as stimuli. The experimental evidence shows that the aperture problem is solved in area MT and several theories have been proposed to explain the process of integration of local motion information in solving the aperture problem in area MT (Movshon et al. 1985, Pack and Born 2001, Bradley and Goyal 2008).

**Pattern motion detection in response to plaids**

Plaids result from the sum of two sinusoidal drifting gratings moving in different directions (Figure 3c). Humans perceive a unified single-pattern movement, which differs from the directions of the components or in some cases their vector sum/average (Figure 3c, green vs. yellow arrows) (Adelson and Movshon 1982). V1 neurons (and some MT cells) respond to the directions of the separate components (yellow arrows). MT pattern cells are selective for the overall motion of plaid patterns (Figure 3c, green arrow), matching human perception (Movshon et al. 1985). Extensive research has been conducted using plaid stimuli to better understand how MT neurons process pattern motion. Plaid patterns contain "extrinsic terminators", which are distinct image features formed at the crossing points of the bars in the gratings. They are called extrinsic because they are formed not by the features of one bar but rather by the crossing of two bars. Therefore, they are extrinsic to each bar's image structure.

**Tracking 2D moving features in response to tilted bars**

The previous examples emphasized how neurons track 1D features (drifting contours). Other research studies have emphasized the tracking of 2D moving features, as occur at the ends of bars, called "intrinsic terminators" (Figure 3b). Terminators in general, both intrinsic and extrinsic, are essential features for motion detection as they represent unambiguous motion information due to their two-dimensional structure. In the case of the aperture problem, revealing the intrinsic terminators of the stimulus immediately breaks the ambiguity in motion detection. There are some neurons in V1 known as end-stopped neurons, which only respond to the endpoints of the stimulus (terminators) (Hubel and Wiesel 1965). Several studies have been conducted to investigate these end-stopped neurons in V1 via neurophysiological recordings and computational modelling using the tilted bar stimulus (Pack et al. 2003, Zarei Eskikand et al. 2016). The study by Pack et al. (2003) showed that end-stopping is a dynamic behaviour. In their experiment, they observed that end-stopped neurons initially respond to the long bar but after 20-30 ms these neurons respond preferentially to the end-points of the bar. The study by Zarei Eskikand et al. (2016) modelled the dynamics of end-stopping feature and its dynamic behaviour through the lateral inhibition between neighbouring neurons.

**Studying local vs. global motion using the barber's pole illusion**

An effective stimulus to study how local motion is integrated to generate a unified global motion perception is the barber's pole illusion. It exemplifies a class of stimuli that generate a unified perception of motion, despite containing a mixture of ambiguous 1D motion cues (oblique contours) and 2D features (bar terminators) that are moving in different directions. For example, the barber's pole shown in Figure 3d is perceived to have predominantly upward motion (green arrow) consistent with the direction of the terminators formed on the long vertical edges (red arrow). This perception arises despite the presence of actual motion to the right (yellow right arrow) on the short horizontal edges and the oblique contour motion (up and right yellow arrows) formed as a local motion perpendicular to the edge of the bars. Consistent with the perception of these stimuli, Pack et al. (2004) showed that many MT neurons can selectively integrate information about the motion of terminators while ignoring ambiguous information about the motion of 1D features (Pack et al. 2004).

**Studying the chopstick illusion and associated neurophysiological findings**

The end of a moving bar is an unambiguous 2D motion cue, an intrinsic terminator. However, when a bar is occluded by another object, the terminators thus formed become extrinsic and their perceived motion may be very different to that of the bar. Figure 3e shows the chopstick illusion, in which two tilted bars are moving right and left (horizontal red arrows) but the motion of the extrinsic terminator (at the X-junction) is perpendicular to this axis (upward yellow arrow). Many neurons in MT are able to distinguish the motion of intrinsic from extrinsic terminators, integrating the correct intrinsic cues while discounting the extrinsic cues (Sharpee et al. 2006). For instance, in Figure 3e, a cell might report rightward motion for the black bar, which corresponds to the direction of the intrinsic terminator.

**Center-surround interaction in response to drifting boxes**

The example of the crossed chopstick illustrates a general principle of visual motion tracking that requires two opposing processes: integration and segmentation. Integration combines motion cues belonging to the same object; segmentation distinguishes between the motion of different objects (Braddick 1993). A prominent theory regarding this distinction involves the detection of motion-similarity for integration versus motion-contrast for segmentation. This mechanism utilizes the known center-surround structure of RFs in the visual system to incorporate contextual information across space. This RF structure allows stimuli in the center to drive responses directly, while stimuli in the surround only modulate responses without evoking spikes. Integration neurons have a facilitatory surround with the same motion preference as the center; segmentation neurons have antagonistic motion preferences (Braddick 1993).

Huang et al. (2008) studied this dichotomy using drifting boxes (Figure 3f) that cover both the center and surround of MT neuron RF (solid vs. dotted blue circles). Many individual MT neurons can shift between integration and segregation in a dynamic stimulus-dependent way. When the motion cues in the RF center are weak due to the low level of contrast or ambiguous because of the aperture problem (1D contours, presented in Figure 3f), the effect of the surround is faciliatory and integrative. However, when the motion cues are unambiguous and strong (2D in the terminators), the effect of the surround is antagonistic (segmentation). Figure 3f shows a drifting box which is moving in the up-right direction. The motion at the edges of the box is ambiguous because of the aperture problem. The experiment shows that MT neurons with the center of RF on the edge of the bar have facilitatory surround (Huang et al. 2008).

However, the surround of the MT neurons switches to antagonistic surround at the terminators when there is no ambiguity

## 2. Computational models

There are many theoretical ways in which a motion signal can be extracted from a moving image. Some models have been developed that have no relationship with known biological systems. Conversely, several models have been created that deliberately attempt to use the building blocks found in natural visual systems as the basis for motion computations. We will describe both purely mathematical and biological-based models to compare and contrast the mechanisms.

### 2.1. Models for local motion processing

**Correlation-based models**

Many existing models of motion perception have their origins in a model proposed by Reichardt (1961), which was modified later by Van Santen and Sperling (1985). This model was guided by observations made during behavioral experiments on walking beetles. Borst, (2007) observed how beetles compensated for their self-motion perception noting that the beetles tended to follow the movement of the visual surround in response to their mistaken perception of self-motion in the opposite direction. The Reichardt detector calculates the direction of image motion using the correlation between the outputs of two receptors separated by a physical distance, i.e., the delayed output of one receptor is multiplied by the output of the other receptor (Van Santen and Sperling 1985). If the time taken for an object to move between the non-delayed signal and the delayed signal is the same as the temporal delay assumed in the model, the correlation is maximal, leading to a maximum response. The direction of motion associated with this maximum correlation is called the "preferred direction". As a result, the Reichardt detector is not only motion selective but also direction selective. The model uses the results of the correlation as the main output. The Reichardt detector takes inspiration from the beetle's compensation behavior. It calculates the direction of image motion using the correlation between the outputs of two receptors separated by a physical distance. In the beetle's case, the two receptors could be analogous to the two eyes, each seeing a slightly different perspective of the visual surround due to their spatial separation. The correlation model was very important in revealing the importance of nonlinear processing in motion detection. Experiments on insect visual systems have gone on to strongly suggest that correlation-like motion detectors exist in insect visual systems (Ibbotson and Goodman 1990, Borst 2007).

**Gradient motion detectors**

In this model, the motion signals from retina are defined as a luminance level, $I$, which depends on time, $t$, and spatial location, $x$. Gradient motion detectors compute the velocity of the motion at each point by the temporal derivative of the luminance function $\partial I(x,t)/\partial t$ divided by the spatial derivative of the luminance $\partial I(x,t)/\partial x$. The motion computation in this model is based on the fact that the rate of temporal variation in the luminance because of motion can be computed by dot multiplication of the rate of spatial variation and the motion's velocity, $v$ (i.e. $\frac{\partial I(x,t)}{\partial t} = \frac{\partial I(x,t)}{\partial x}.v$) (Borst 2007). It has been suggested that the sustained signals observed in "sustained retinal ganglion cells" indicate the spatial variation in the luminance, while the transient signals in "transient retinal ganglion cells" indicate the temporal variation in the luminance (Mather 1984). The responses of transient cells depend on the temporal frequency of the stimulus . These cell types suggest that the building blocks for gradient motion detectors are present in the biological system, making this type of model plausible in biology.

The problem with gradient motion detectors is that the model depends on observing the image gradient at all times, which is not always possible. Therefore, the output of the model is not defined in homogeneous regions of the image where the spatial derivative is zero. In addition, in cases where the spatial derivative is very small, a small amount of noise in the system will lead to a large increment in the final output, causing this model to fail for cases where the signal to noise ratio is small (Borst 2007).

**Spatiotemporal motion energy models**

Spatiotemporal motion energy models estimate the direction and velocity of moving images based on the outputs of spatiotemporal filters. Energy models extract motion information using oriented Gabor filters, which resemble the functions of the oriented RFs found in V1 (Adelson and Bergen 1985). While the internal structures of energy models differ from those of the correlation model, the processing generated is nearly equivalent to the final output. Gabor filters are defined by the multiplication of sinusoidal waves and a Gaussian function. The oriented spatiotemporal filters used for modeling complex neurons in V1 respond strongly to motion in their preferred direction and have zero activity when there is no movement. This method provides a good fit to the temporal dynamics of V1 responses (Adelson and Bergen 1985). The Gabor function has been used extensively to model the receptive fields of neurons and it is the best way to mathematically describe the reduced response with distance from the centre of the blob.

The structure of energy models is a close fit to the multi-filter structures proposed to explain complex cell behavior in cat and monkey cortex, i.e., at least two Gabor filters that are phase shifted relative to each other and combined in a nonlinear way (Adelson and Bergen 1985). Interestingly, recent work in cat cortex has used the nonlinear input model to reveal for the first time both the spatial filters and the nonlinearities that underly the behavior of V1 complex cells (Almasi et al. 2020). These studies have confirmed the existence of most of the theoretical components of complex cells, including the essential elements that would make motion detection possible through computations similar to those in the energy model.

**Modeling systems tuned to the speed of motion**

Neurophysiological findings by Perrone and Thiele (2001) revealed that a large proportion of MT neurons are speed tuned, where the neurons respond maximally to edges moving at particular speeds within their receptive fields. Speed tuning in natural visual systems has been noted in multiple species including insects such as flies (Collett and Land 1978, Nordström and O'Carroll 2006) and bees (Srinivasan et al. 1991, Ibbotson et al. 2017). Perrone and Thiele (2002) proposed a method, using motion energy filters as building blocks, to replicate the speed tuning of neurons in area MT. In their method, they combine two spatiotemporal filters representing two different types of transient and sustained V1 neurons in primates (Perrone and Thiele 2002). The sustained and transient V1 neurons have different temporal frequency response profiles. Sustained V1 neurons respond best to static stimuli and their temporal frequency response is that of a low pass filter, while transient V1 neurons respond better to moving stimuli and have a band pass temporal frequency profile (Foster et al. 1985) (Figure 4a). Combining these two temporal functions by applying an absolute and inversion operation on their difference (i.e., $(|\log T(f) - \log S(f)| + 0.75)^{-1}$, where $T(f)$ is the temporal function of the transient neurons and $S(f)$ is the temporal function of the sustained neurons) will result in a narrowly tuned temporal function (Figure 4b). Perrone and Thiele (2002) incorporated this feature in their model to replicate the speed tuning of MT neurons. The data by Nover et al. (2005) shows that the scheme of speed coding in MT is coded logarithmically and it is scale invariant.

a.

b.

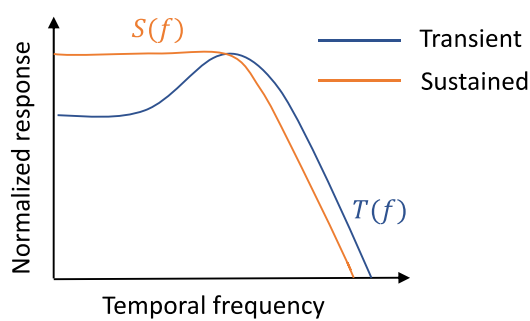
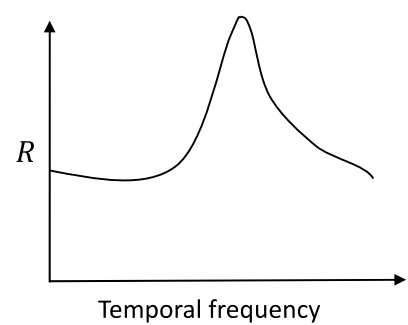

Figure 4. (a) Transient V1 neurons have a band-pass temporal frequency profile while sustained V1 neurons have a low-pass temporal frequency profile. (b) Operating an inversion and absolute operation on the difference between the sustained and transient temporal function, $R = \frac{1}{|\log T(f) - \log S(f)| + 0.75}$, will result in a narrowly tuned response to the speed of the stimulus. 0.75 is a constant that determines the width of the tuning curve. Axes are in log space.

The data recorded by Priebe et al. (2003) shows that there is variability in the dependence of the speed tuning of the MT neurons on the spatial frequency of the stimulus, and only 25% of the MT neurons maintain their speed tuning with changes in spatial frequency. This suggests that the tilted spatiotemporal response profile suggested by Perrone and Thiele (2001) is a necessary but not sufficient condition to indicate the speed tuning of the neurons. It is important to note that these are abstract models of the functions of the neurons within the network and do not necessarily represent the actual synaptic connections between different types of neurons.

There exist other alternative methods suggested to simulate the speed tuning of MT neurons. The model by Chey et al. (1998) uses a multiscale filter approach to extract the speed of the stimulus. Different filter widths in this model give sensitivity to different ranges of speeds. In this model a "scale-proportionate threshold" is applied to prevent the larger scales from having the highest activity at all times, even when the speed of the stimulus is low. Although, the scale-proportionate threshold is a reasonable approach to solve this problem, it causes difficulty in the detection of the stimulus at low levels of contrast. Perrone (2012) deals with this problem by adding an inhibition between different units for speed detection. This inhibition limits the output of each unit that is responding maximally to the stimulus. For example, the response of an MT unit responding to a stimulus because it is covering a large area but is not tuned to the speed will be suppressed by another unit with larger response, which covers the same area of the receptive field but is tuned to the speed of the stimulus (Perrone 2012).

## 2.2. Models to deal with the aperture problem

The models outlined in Section 2.1 are designed to detect motion locally, using only the information in one small RF location. In this section, we expand on this local processing to the next tier of motion coding, which seeks to solve some of the problems generated by the need to initially break down the image into small segments (i.e., as caused by small RFs). The motion information extracted from these small segments by local motion processing methods suffers from these problems in the same way as most neurons in the primary visual cortex (V1). Therefore, these initial motion signals need to be processed further to recover the true global direction of motion.

Neurons in V1 in the primate brain show a pronounced tuning for oriented features such as edges. V1 neurons can only measure local motion signals, i.e., the component of the motion that moves orthogonal to the orientation of an edge (Figure 5, up-right yellow arrow). Any component of the motion parallel to the edge is not detectable by complex neurons because of the invariance of the contrast in this direction (Figure 5, down-right yellow arrow). Measuring only one component of the motion results in an ambiguity of the direction of motion, called the "aperture problem". When viewed through a narrow V1 receptive field, the edge of a moving bar only appears to move perpendicular to the edge of the bar (Figure 5) (Pack et al. 2003, Bradley and Goyal 2008). The experimental evidence shows that the aperture problem is solved in MT and several theories have been proposed to explain the process of integration of local motion information in solving the aperture problem in MT (Bradley and Goyal 2008).

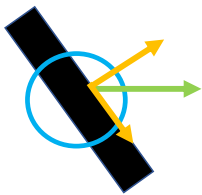

Figure 5. A tilted bar stimulus and the aperture problem. The blue circle represents the RF of a neuron. Yellow arrows are the local component of the motion, perpendicular to the edge and in parallel to the edge of the bar,

green arrow is the actual direction of motion of the bar. Looking through the aperture (resembling the RF), it appears that the bar is moving in an up-right direction.

**Vector average method**

There are several motion detection models proposed to deal with the ambiguous motion signals resulting from the aperture problem. One of the simplest proposed methods in this category is known as the vector average method (Mingolla et al. 1992). This method calculates the average value of local velocities to obtain the global velocity of the object in a larger spatial range. Although the resulting value is highly correlated with the object's global direction and speed, it is a poor estimator of actual velocity because the velocity of the vector average method is highly dependent on the shape of the object (Figure 6). Therefore, objects that have different shapes and the same speed will have different vector averages (Bradley and Goyal 2008).

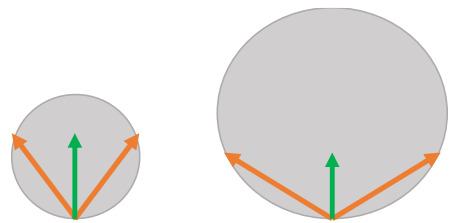

Figure 6. Velocity diagram describing the misrepresentation of global motion using the vector average method. Although the local motion of the object on the right has a larger magnitude (orange arrows) (representing a higher speed), the estimated global motion (green arrow) using the vector average method is lower than for the object on the left-hand side (Bradley and Goyal, 2008).

**Intersection of constraints model**

Another simple method to overcome the aperture problem is known as the intersection of constraints model. This method computes the global motion by combining the local motions of the stimulus. Any local motion vector has a perpendicular component that, when drawing a perpendicular line from the tips of the local motion vectors of two edges of a stimulus and finding the intersections of these lines, will result in the computation of the global motion of the stimulus (Figure 7) (Adelson and Movshon 1982). The main drawback of the intersection of constraints method is that it must assume that motion components belonging to the same object are already grouped, so it cannot distinguish motion of multiple moving stimuli in the field of view.

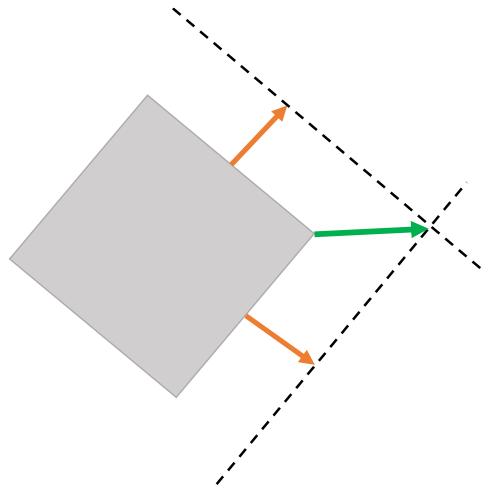

Figure 7. Diagram illustrating the intersection of constraints method. The one-dimensional local motion component (orange arrow) represents an ambiguous estimation of the global direction (green arrow). To estimate the global motion signal using two of the component motions, a line (black dotted lines) is drawn perpendicular to the tips of the component motions.

All of the methods that we have discussed so far represent the linear features of V1 neurons. However, the responses of the neurons are not entirely linear. In addition, to achieve a more biologically plausible model, there should also be some mechanisms embedded in the model to implement the saturation of the responses of the neurons. The following methods discuss models that incorporate nonlinearities into the responses of the neurons as well as their linear characteristics in solving the aperture problem.

**Linear-nonlinear model**

To solve the aperture problem, Simoncelli and Heeger (1998) proposed a linear-nonlinear model based on the intersection of constraints method. The linear-nonlinear model has two stages, which correspond to areas V1 and MT in the primate brain.

The stimulus is passed through the first stage of the model, which comprises oriented spatiotemporal filters representing models of V1 neurons. The output of the filters is rectified and then a nonlinear normalization followed by a spatial pooling is implemented on the resulting activities of the neurons.

The technique applied in this model is more readily explained in the frequency domain. The Fourier transform of a Gabor function is a pair of fuzzy blobs in the frequency domain in which their density decreases with distance from the center of the blob. An image can also be described in the frequency domain with a spectrum reflecting the image intensity at various frequencies. A simple cell responds well when the spectrum plane of the image (stimulus) passes through the center of the blobs representing the Fourier transform of the Gabor functions, which describe the receptive fields of the neurons (Figure 8a). The center of a blob is called the center frequency. The aperture problem is defined in the frequency domain by the fact that there are multiple planes passing through the center of the blob with different orientations that can generate the maximum response of the neuron (Figure 8b). The linear-nonlinear model solves the aperture problem at the next stage of the model where a particular set of V1 complex neurons that have center frequencies on the same plane are summed to generate the responses of pattern MT neurons. The orientation of this plane in the frequency domain determines the preferred velocity of a pattern MT neuron (Simoncelli and Heeger 1998). This pooling mechanism over the neurons with center frequencies on the same plane resolves the aperture problem.

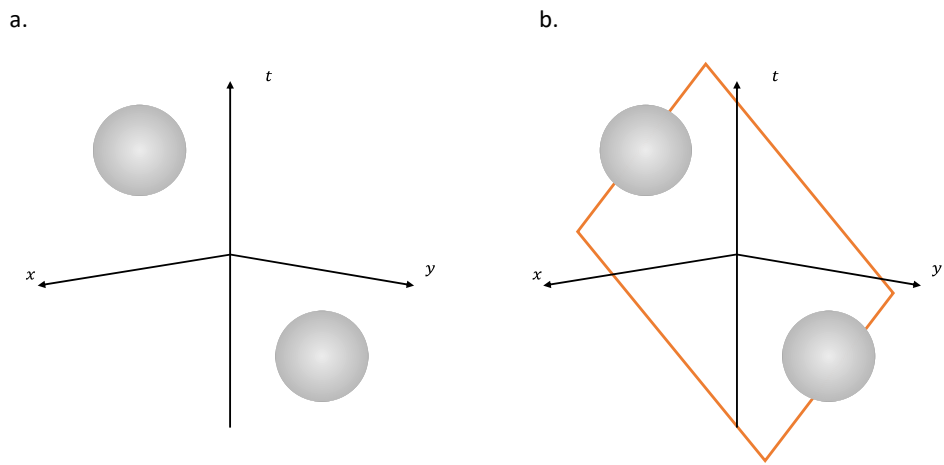

Figure 8. The spectrum of the image is computed by applying a Fourier transform to the image, which results in three-dimensional frequency components ($\omega_x, \omega_y$ and $\omega_t$) for an image in the spatial domain of $x$ and $y$, plus the temporal domain of $t$. (a) The fuzzy blobs, which are symmetrically located about the origin, represent the Fourier transform of the Gabor function modeling the selectivity of a V1 neuron. (b) The response of the modelled neuron to a stimulus will be maximized when the image spectrum passes through the center of the fuzzy blobs.

Rust et al. (2006) proposed an elaborated version of the linear-nonlinear model to capture the main characteristics of the responses of MT neurons. In this model, a specific type of tuned normalization is applied to the output of the V1 neurons, which resembles surround suppression. The responses of the individual neurons are inhibited in proportion to their own responses as a result of the tuned component of the normalization pool.

**Models of end-stopped neurons and surround suppression**

There are some neurons in V1 known as end-stopped neurons, which only respond to the endpoints of the stimulus (terminators) (Hubel and Wiesel 1965). In the case of the aperture problem, revealing the endpoints of the stimulus immediately breaks the ambiguity in motion detection. The study by Pack et al. (2003) has shown that end-stopping is a dynamic behaviour. To deal with the aperture problem, there are models that rely on the specific characteristics of neurons that have suppressive surrounds e.g., end-stopping feature in some V1 neurons. The model by Zarei Eskikand et al. (2016) uses the unambiguous motion information at the terminators to resolve the ambiguous motion information that results from the aperture problem. The unambiguous motion information generated by the end-stopped neurons is propagated to the other regions in MT. Therefore, MT neurons respond to the correct direction of motion after a delay, which resembles the neurophysiological findings by Pack and Born (2001). The end-stopping feature in Zarei Eskikand et al.'s (2016) model is created using lateral inhibitory connections between aligned neighboring V1 neurons selective to the same direction.

There are other models that rely on end-stopping to model the responses of MT neurons. The model of Tsui et al. (2010) also uses motion energy filters that receive modulations from nearby neurons with similar directional preferences. In this case, end-stopping is modelled by the interaction of the center and surround units. To generate the model of surround inhibition, the model sums the outputs of V1 neurons (motion-energy filters) placed at either side of the center units. However, instead of the summation of the raw outputs of the neurons, the envelope of the temporal response is applied. In this way, the neuron is maximally inhibited when the surround units are activated simultaneously. The responses of the MT neurons are computed by applying a nonlinear summation on the outputs of the end-stopped neurons (Tsui et al. 2010).

The tuned nonlinear normalization in the model proposed by Rust et al. (2006) eliminates the ambiguous motion signals resulting from the aperture problem, which is similar to the role of end-stopped neurons in the model proposed by Tsui et al. (2010). However, in contrast to this model, the spatial structure of the normalization pool is not considered in the model proposed by Rust et al. (2006), where the normalization pool is implemented by dividing the linear response of each neuron by a value that is proportional to the summation of the activity of the neurons in the neighborhood. To simulate the responses of the neurons to a tilted bar as a stimulus, the spatial arrangement of the inhibitory units should be considered. The selectivity of the neurons in the model to the endpoints of the stimulus, or any other spatial features, also requires a model that can determine spatial structure. Even though the linear-nonlinear method in Rust et al. (2006) is robust at dealing with complex motion in natural scenes compared to the model of end-stopped neurons in Zarei Eskikand et al. (2016), the linear-nonlinear model lacks the temporal dynamics to describe the dynamic integration of the activities of the MT neurons. This is necessary in the description of the responses of MT neurons to plaid stimuli and for explaining the associated temporal delay in perception of pattern motion.

**Models simulating propagation of motion information**

There are several models that propose recovering motion information at the ambiguous locations in the image through the propagation of activity of the neurons from one region to another within MT. These networks use recurrent interconnections or feedback connections to propagate the motion information (Lidén and Pack 1999, Bayerl and Neumann 2004).

Pack and Born (2001) experimentally recorded the responses of MT neurons to moving stimuli over time. They observed that MT neurons respond to ambiguous motion information at the onset of the stimulus, and the neurons overcome the ambiguity of the motion representation after a temporal delay of around 200ms. This suggests the possibility of propagation of the unambiguous motion information from the terminators at the MT level, which requires a certain amount of time.

In another model proposed by Liden and Pack (1999), the aperture problem is solved by the propagation of motion information between MT neurons. There is no strategy in this suggested model to discriminate ambiguous and unambiguous motion information. Therefore, the neighboring neurons representing the ambiguous motion information that are active simultaneously over a larger area dominate the other neurons. This results in the suppression of the unambiguous motion information resulting from the response to the endpoints of the stimulus.

In contrast, the model proposed by Zarei Eskikand et al. (2016), which has a similar scheme, applies a model of end-stopped neurons in its input to differentiate the unambiguous motion information generated at the terminators of the stimulus. The excitatory and inhibitory interconnections between MT neurons results in the propagation of unambiguous motion information at the terminators to the other neurons in the network. The lateral propagation of activities in the model replicates the neurophysiological findings on the temporal dynamics of the real MT neurons (Pack et al. 2003).

**Models applying feedback connections**

Several models have suggested methods for estimating the correct direction of motion by strengthening unambiguous motion signals relative to ambiguous signals via feedback (Grossberg et al. 2001, Bayerl and Neumann 2004). In these models, the propagation of motion information requires a second stage, called the motion grouping network (Grossberg et al. 2001). In these models, neurons with similar direction preferences are selected as the winning neurons and they suppress neurons with other directional preferences using feedback connections. Therefore, for the propagation of motion information in these models, a model of the MST area, which is one step beyond MT, is incorporated (Grossberg et al. 2001).

A model proposed by Beck and Neumann (2011) uses a similar approach of propagation of motion information. However, in this model, a combination of feedback connections from MT neurons and lateral inhibition is used to suppress the activity of the neurons with ambiguous motion information. According to this model, the ambiguous perception of motion by neurons at the early stages of motion processing (i.e.V1 neurons) is resolved through the feedback connections from MT neurons to V1. MT neurons have larger receptive fields and as a result they carry less ambiguous motion information compared to V1 neurons. Therefore, a feedback connection from MT to V1 assists in resolving the ambiguity in the perceived motion information in V1 (Beck and Neumann 2011). The temporal evolution of the response of the MT neurons has been observed through the experiments of Pack and Born (2001). However, there is no experimental evidence to show that standard complex V1 neurons follow the same temporal pattern. The model by Beck and Neumann (2011) also applied a model of end-stopped neurons to disambiguate the local motion information. This accords to the neurophysiological evidence from Pack et al. (2004), who showed that MT neurons are more influenced by the unambiguous motion information of the endpoints.

## 2.3. Models describing the interaction of form and motion information

There is experimental evidence that the division of form and motion processing into the ventral and dorsal streams of the primate brain is not absolute; form information may contribute to the processing of motion information in the dorsal pathway (Krekelberg et al. 2003). There are several computational models that propose the involvement of form information in the processing of image motion. Models proposed by Grossberg (1994) show how motion and form information cooperate to achieve figure-ground segregation, which is identifying an object from its background. These models are based on the interaction of three systems: figure-ground segregation to discriminate the occluders from the occluding object properties, a motion processing stream, and a system to integrate motion and form information.

Form information in these models is captured through the model of bipole neurons, which have two separated RFs. There is neurophysiological evidence illustrating the existence of bipole neurons, which have multiple neuronal RFs. These RFs seem to be connecting different sections of a discontinuous stimulus (illusory contours) that appear to belong to the same object (Von der Heydt et al. 1984). The main role of form information in the model proposed by Grossberg (1994) is connecting the illusory contours, formed as the result of occlusion with another object, to form a single coherent moving feature to guide the integration processes of MT neurons (Grossberg 1994). The term "illusory contours" in these models refers to the contours formed when an object passes in front of another object, thus occluding the contours of the background object. It is not known whether the global form of the stimulus is perceived in areas MT and MST or if this perception is generated in higher levels of cortex.

In contrast to the models that apply form information mainly to achieve a unified image of the stimulus in the presence of occluders (Grossberg 1994), Zarei Eskikand et al. (2018)

proposed a model that suggests the direct involvement of form information to generate an accurate perception of stimulus motion. This is performed by compensating for the effect of extrinsic terminators using form-related information generated early in the visual system. Extrinsic terminators are formed as a result of overlapping stimuli in the visual fields, which transmit conflicting motion information (Zarei Eskikand et al. 2018, Zarei Eskikand et al. 2020).

In the model proposed by Raudies and Neumann (2010), the motion information is processed through a hierarchical model of different areas in the dorsal pathway (i.e., areas V1, MT, and MST), and the form information is processed in areas V2 and V4 to generate modulating feedback that arises from spatial attention signals. The interaction of form and motion information occurs through modulating feedback via V4 neurons. In these models, feedback connections from higher stages with larger receptive fields result in evolution of the initial rough estimation of motion in V1, which ultimately leads to the correct estimation of motion. This model implements feedback connections from MT to propagate unambiguous activities from MT to V1 neurons, suggesting additional processing that evolves over time.

## 2.4. Models to explain pattern motion processing of MT neurons

Several models have been proposed to explain the mechanism of pattern motion selectivity of MT neurons. Most of these models suggest a hierarchical relationship between component and pattern selective MT neurons (Rust et al. 2006). This theory of pattern motion perception, which relies on the summation of component motions of the gratings, is very elegant and corresponds to the properties of certain parts of the cortical visual pathways (Rodman and Albright 1989). However, the theory is developed largely around the use of plaids. Plaid patterns are a rather artificial concept. Where two gratings cross in a plaid, blobs are formed that move in the pattern direction. In natural scenes, when oriented structures such as tree trunks move relative to each other it is usual for the ends of these oriented structures (intrinsic terminators) to be readily visible, such as the ends of branches. In human perception, revealing the ends of moving gratings immediately breaks the illusion of the direction of pattern motion: the true direction of motion is perceived.

Based on this simple observation, Zarei Eskikand et al. (2019) developed a model that emphasizes the role of terminators in the process of motion information. According to this theory, the visual system extracts the direction of motion of terminators as its best estimate of the direction of object motion. If those terminators are not available within a cell's RF (e.g., with plaids centered on a relatively small RF), MT falls back upon the next best method, which is to extract the direction of motion based on the direction of oriented edges or the extrinsic terminators formed at the intersections of the gratings. Based on the specific combination of inputs from V1, it may be that some MT neurons emphasize more of the local motion information as the result of extrinsic terminators and have been labelled pattern cells, while others emphasize the motions of the edges of the gratings leading to intermediate or component selective cells.

This theory is consistent with neurophysiological findings (Kumbhani et al., 2015; Majaj et al., 2007), which showed the spatial and temporal limits of the pattern motion selectivity of MT neurons. According to their observations, for pattern MT neurons to respond to the pattern motion of the plaids, the gratings must appear simultaneously in a limited temporal and spatial range. This experiment shows that pattern MT neurons respond to the motion of the individual gratings when presented with pseudo-plaids, which are two independent gratings that appear simultaneously in two parts of the same receptive field (Fig. 9a, top). Pattern MT neurons do not integrate the component motion of the individual gratings in this case where there is no overlap between the components of the gratings. This is consistent with the theory that processing of extrinsic terminators might play a key factor in the responses of the neurons to the pattern motion of the stimulus.

Several models, described below, attempted to explain the neurophysiological findings by Majaj et al. (2007). In the model of Beck and Neumann (2011), MT neurons respond to the motion of the individual non-overlapped bars in their RFs without integrating their local motion signals as a single pattern motion. This conclusion has been disputed as the applied stimuli used to test this model are different to those used by Majaj et al. (2007). Beck and Neumann applied a stimulus where the intrinsic terminators of the bars were visible within the RFs of the neurons, which is not equivalent to the pseudo-plaids used by Majaj et al. (2007) (Figure 9b). In this case, MT neurons will represent the motion of the intrinsic terminators (e.g., motion of the ends of the bars) but the motion of the overlapping bars (extrinsic terminators) cannot represent the pattern motion observed in the plaids. In the case where overlapping bars are the stimulus, the MT neurons should respond to the motion of the individual bars by overcoming the conflicting motion at the extrinsic terminators.

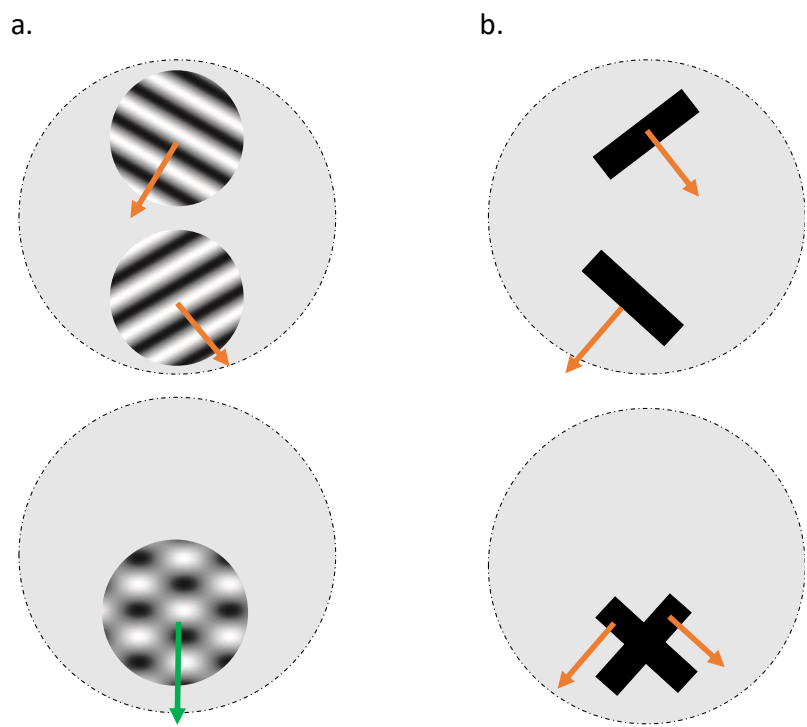

Figure 9. Pseudo-plaids are two individual gratings that appear simultaneously in a single receptive field of a neuron (shown as the dotted circle (a, upper) and a plaid stimulus formed by the integration of the individual

gratings, a, lower). Two separate bars moving in different directions simultaneously within a single receptive field of the neuron (b, upper) and two overlapped bars moving in different directions (b, lower).

Perrone and Krauzlis (2008) also proposed a model to explain the results of the neurophysiological experiments by Majaj et al. (2007). This model shows that the response of the neurons to the component motions of the pseudo-plaids results from subregions in the RFs of the MT neurons. Therefore, when the gratings of a plaid are spatially separated and they are stimulating in different regions within the RFs, the pattern MT neurons do not integrate the responses of the gratings and respond to the component motions instead. The model by Perrone and Krauzlis (2008) concludes that pattern motion detection in MT neurons occurs in the subregions of the RFs of the MT neurons.

The model proposed by Zarei Eskikand et al. (2019) has a different view of this neurophysiological experiment than the model of Perrone and Krauzlis (2008), even though this model does not cover the velocity preference of MT neurons, in contrast to Perrone and Krauzlis (2008). Zarei Eskikand et al.'s model theorizes that the strong component response to the pseudo-plaids can result from the absence of extrinsic terminators in these gratings. The spatially separated gratings of a plaid do not have extrinsic terminators, which appears to be the key factor in the pattern motion computation of the MT neurons in this model. Therefore, when the gratings overlap, the extrinsic terminators are shaped at the intersections of the gratings and, consequently, the neurons respond to the pattern motion of the stimulus. The spatial and temporal limits on the detection of the pattern motion, which is satisfied by grating overlap, highlights the role of terminators. To evaluate which hypothesis better explains the process of pattern motion detection, experimental frameworks derived from the computational models are proposed below.

Our first hypothesis, driven by the model proposed by Zarei Eskikand et al. (2019), shows the role of stimulus terminators in calculating pattern motion information by neurons in MT. To investigate the effect of intrinsic terminators (e.g., the ends of bars) on the pattern or component selectivity of MT neurons, the activities of the neurons would be measured in response to a stimulus that has two sets of gratings with visible intrinsic terminators moving perpendicular to each other (Figure 10a). According to the theory, the hypothesis is that pattern MT neurons should respond to the pattern motion at motion onset but, after a short temporal delay, the responses should become dominated by the component motions of the individual gratings (Figure 10b). In the second part of the experiment, the terminators of the stimuli should be covered using a mask (occluder) (Figure 10c). According to the theory, with the occluder hiding the intrinsic terminators, pattern MT neurons should be selective only for the pattern motion (Figure 10d). This simple experiment highlights the role of intrinsic terminators in the pattern motion detection of MT neurons, and the switch from a preference for component motion detection to pattern motion detection. Such an observation would reveal the role of intrinsic terminators in the determination of pattern or component selectivity of MT neurons.

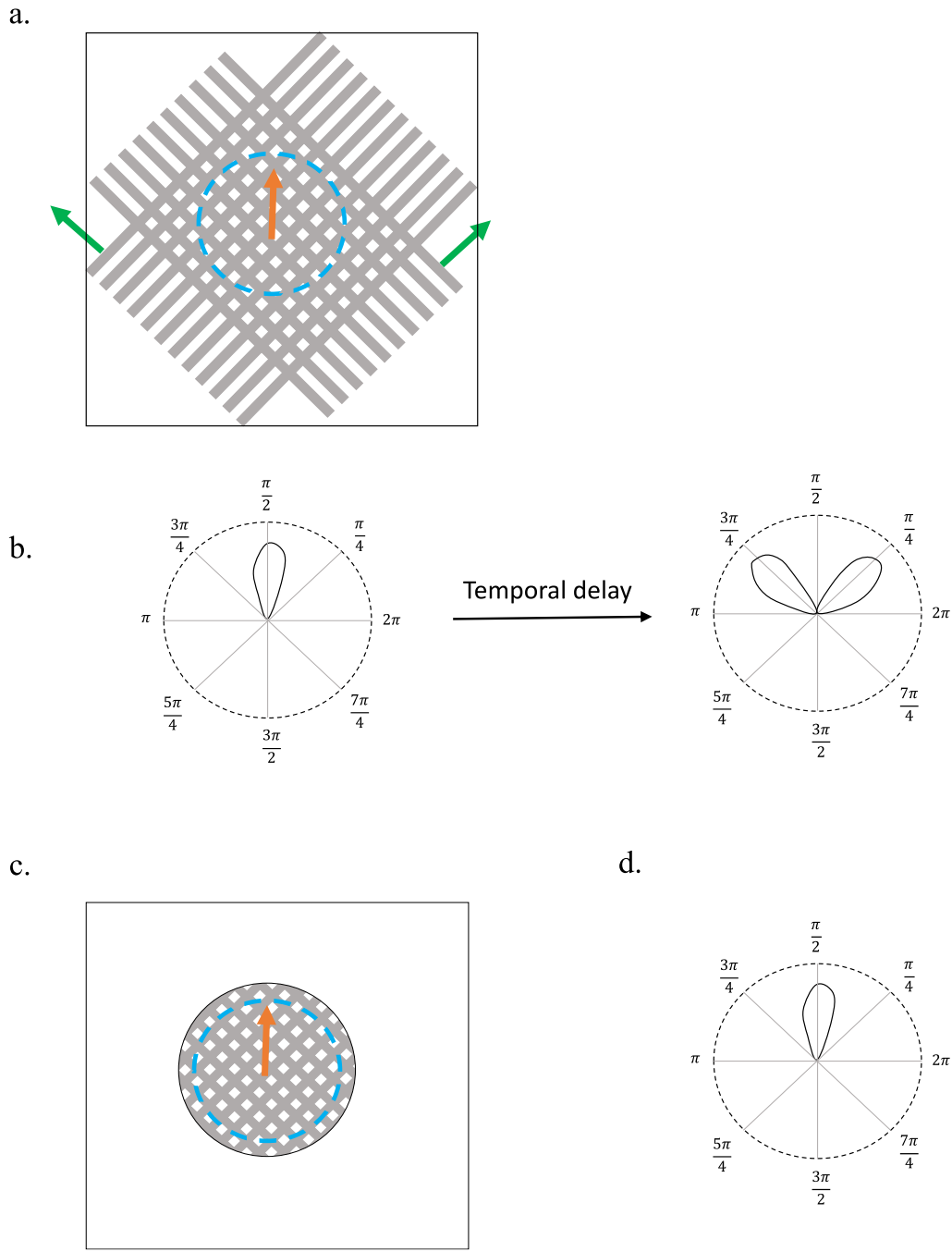

Figure 10. The stimuli for the proposed experiment and the prediction of the computational model derived by Zarei Eskikand et al. (2019). a) Individual gratings moving simultaneously in two different directions, indicated by the green arrows (up-left and up-right). The blue circle shows the receptive field of the MT neuron. The direction of the perceived pattern motion is shown by the orange arrow. b) The predicted response of the MT neurons according to the theory, which predicts that the neurons respond to the pattern motion at the beginning of the stimulation period if terminators are available in the extra-classical receptive field regions of the neuron. The theory then suggests that due to interactions within a larger population of MT neurons, the tuning of the recorded

cell switches to a selectivity for the component motions after a temporal delay. c) Individual gratings moving in different directions (up-left and up-right) with their terminators occluded: as a result, it appears that the stimulus is moving in the upward direction shown by the yellow arrow. The classical receptive field of the neuron is shown by the blue circle. d) According to the theory, the pattern MT neurons respond to the pattern motion when the terminators of the stimulus are covered using an occluder.

To continue the exploration of the temporal dynamics of MT neurons, other experiments could be performed. Experiments by Smith et al. (2005), using plaid patterns without intrinsic terminators, recorded a considerable temporal delay of 50-70 ms in the detection of pattern motion by MT neurons compared to component motion. In this scenario, with only the information from extrinsic terminators within the plaid available (and no intrinsic terminators), MT neurons were found to be initially selective to the components of the plaids. Only after a delay did the cells become selective for pattern motion. According to the hypothesis suggested by Zarei Eskikand et al. (2019), the propagation of the activity from the extrinsic terminators within the plaids results in a temporal delay in the perception of pattern motion, while component-selective neurons have a consistent response over time. In accord with the theory, the consistency of the direction selectivity of component MT neurons has been observed (Smith et al. 2005).

To test this hypothesis, the activity of MT neurons could be examined in response to two plaids, made from low and high spatial frequency gratings, respectively (Figure 11). Importantly, the high spatial frequency plaid has a higher density of extrinsic terminators, with smaller spatial separation between terminators. If the temporal delay in the detection of the pattern motion results from the propagation of the activity of the neurons from the extrinsic terminators, we expect a longer delay in the detection of the pattern motion of plaid (b) (low spatial frequency) compared to plaid (a) (high spatial frequency), as the longer distances between the intersections of the plaid require more prolonged propagation. However, if the detection of the pattern motion is based on the summation of the component motions of the individual gratings, there will not be a significant difference in the temporal dynamics of the pattern motion detection of these plaids.

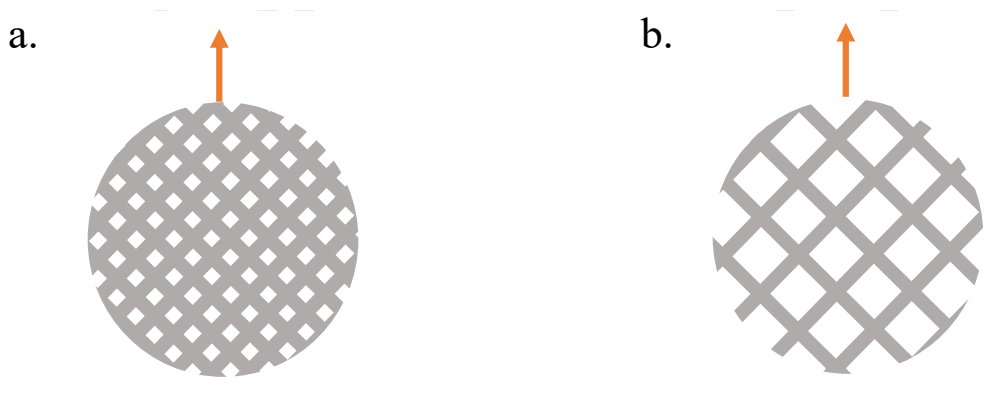

Figure 11. The stimulus suggested to investigate the temporal dynamics of pattern motion selectivity of MT neurons. The plaid in (a) has a higher density of extrinsic terminators due to the higher spatial frequencies of the grating components, compared to (b).

## 4. Conclusion

There is a rich history of modelling the visual system using a combination of experimental and computational techniques. Computational modelling of motion processing in the visual cortex has been successful in replicating several neurophysiological findings to better understand the detailed circuitry of biological visual systems and the role of different types of neurons in those systems. More specifically for motion detection, models have been proposed to explain the circuitry of the neurons in dealing with the aperture problem. These models have driven several hypotheses relating to the perception of motion by V1 and MT neurons in primates, specifically by identifying the role of end-stopped neurons in V1 and surround suppression effects in the detection of motion information. There are several models that explain the interaction of form and motion information and describe the contribution of form information in motion processing across several areas in the visual cortex. Computational modelling studies have also been successful in replicating neurophysiological findings describing the perception of pattern motion in primate MT neurons. There are several hypotheses driven by computational models on the role of MT neurons in the perception of motion. Experiments aiming to test these hypotheses may have a significant impact on our understanding of the processes of pattern motion detection in the visual cortex.

Most of the models proposed to explain the function of MT neurons are not detailed models and they generally apply sophisticated mathematical tools to explain specific behaviour observed in the neurophysiological experiments in an abstract form. This review paper brings together all the relevant experimental evidence related to existing mathematical models, although there is always a knowledge gap where there is not yet exist experimental evidence or the methods are not biologically realistic.

# References

Adelson, E. H. and Bergen, J. R. (1985). Spatiotemporal energy models for the perception of motion. Journal of the Optical Society of America A **2**: 284-299.

Adelson, E. H. and Movshon, J. A. (1982). Phenomenal coherence of moving visual patterns. Nature **300**: 523-525.

Almasi, A., Meffin, H. , Cloherty, S. L. , Wong, Y. , Yunzab M. and Ibbotson, M. R. (2020). Mechanisms of Feature Selectivity and Invariance in Primary Visual Cortex. Cerebral Cortex **30**: 5067-5087.

Bayerl, P. and Neumann, H. (2004). Disambiguating visual motion through contextual feedback modulation. Neural Computation **16**: 2041-2066.

Beck, C. and Neumann, H. (2011). Combining feature selection and integration—a neural model for MT motion selectivity. PLoS One **6**: e21254.

Borst, A. (2007). Correlation versus gradient type motion detectors: the pros and cons. Philosophical Transactions of the Royal Society B: Biological Sciences **362**: 369-374.

Bradley, D. C. and Goyal, M. S. (2008). Velocity computation in the primate visual system. Nature Reviews Neuroscience **9**: 686-695.

Chey, J., Grossberg, S., and Mingolla, E. (1998). Neural dynamics of motion processing and speed discrimination. Vision Research **38**: 2769-2786.

Collett, T. and Land, M. (1978). How hoverflies compute interception courses. Journal of Comparative Physiology **125**: 191-204.

Foster, K., Gaska, J. P. , Nagler, M. and Pollen, D. (1985). Spatial and temporal frequency selectivity of neurones in visual cortical areas V1 and V2 of the macaque monkey. The Journal of Physiology **365**: 331-363.

Fukushima, K. and Kikuchi, M. (1995). Binding of form and motion by selective visual attention: A neural network model. Brain Processes, Theories, and Models. An International Conference in Honor of WS McCulloch 25 Years after his Death: 436-445.

Goodale, M. A. and Milner, A. D. (1992). Separate visual pathways for perception and action. Trends in Neurosciences 15: 20-25.

Grossberg, S. (1994). 3-D vision and figure-ground separation by visual cortex." Perception & Psychophysics 55: 48-121.

Grossberg, S., E. Mingolla and L. Viswanathan (2001). Neural dynamics of motion integration and segmentation within and across apertures." Vision Research 41: 2521-2553.

Gur, M. and Snodderly, D. M. (2007). Direction selectivity in V1 of alert monkeys: evidence for parallel pathways for motion processing. The Journal of Physiology 585: 383-400.

Hubel, D. H. and Wiesel, T. N. (1965). Receptive fields and functional architecture in two nonstriate visual areas (18 and 19) of the cat. Journal of Neurophysiology 28: 229-289.

Ibbotson, M., Y.-S. Hung, H. Meffin, N. Boeddeker and M. Srinivasan (2017). "Neural basis of forward flight control and landing in honeybees." Scientific Reports 7: 1-15.

Ibbotson, M. R. and Goodman, L. J. (1990). Response characteristics of four wide-field motion-sensitive descending interneurones in Apis Melufera. Journal of Experimental Biology 148: 255-279.

Krekelberg, B., Dannenberg, S., Hoffmann, K. P., Bremmer, F. and Ross, J. (2003). Neural correlates of implied motion. Nature 424: 674-677.

Lidén, L. and Pack, C. (1999). The role of terminators and occlusion cues in motion integration and segmentation: a neural network model. Vision Research 39: 3301-3320.

Mather, G. (1984). Luminance change generates apparent movement: Implications for models of directional specificity in the human visual system. Vision Research 24: 1399-1405.

Maunsell, J. and van Essen, D. C. (1983). The connections of the middle temporal visual area (MT) and their relationship to a cortical hierarchy in the macaque monkey. Journal of Neuroscience 3: 2563-2586.

Mingolla, E., Todd J. T. and Norman, J. F. (1992). The perception of globally coherent motion. Vision Research 32: 1015-1031.

Movshon, J. A. and Newsome, W. T. (1996). Visual response properties of striate cortical neurons projecting to area MT in macaque monkeys. Journal of Neuroscience 16: 7733-7741.

Nordström, K. and O'Carroll, D. C. (2006). Small object detection neurons in female hoverflies. Proceedings of the Royal Society B: Biological Sciences 273: 1211-1216.

Pack, C. C., Livingstone, M. S., Duffy, K. R. and Born, R. T. (2003). End-stopping and the aperture problem: two-dimensional motion signals in macaque V1. Neuron 39: 671-680.

Perrone, J. A. (2012). A neural-based code for computing image velocity from small sets of middle temporal (MT/V5) neuron inputs."Journal of Vision 12: 1-1.

Perrone, J. A. and Thiele, A. (2001). Speed skills: measuring the visual speed analyzing properties of primate MT neurons. Nature Neuroscience 4: 526-532.

Perrone, J. A. and Thiele, A. (2002). A model of speed tuning in MT neurons. Vision Research **42**: 1035-1051.

Priebe, N. J., Cassanello, C. R. and Lisberger, S. G. (2003). The neural representation of speed in macaque area MT/V5. Journal of Neuroscience **23**: 5650-5661.

Reichardt, W. (1961). Autocorrelation, a principle for evaluation of sensory information by the central nervous system. Symposium on Principles of Sensory Communication 1959, MIT press**:** 303-317.

Rodman, H. R. and Albright, T. D. (1989). Single-unit analysis of pattern-motion selective properties in the middle temporal visual area (MT). Experimental Brain Research **75**: 53-64.

Rust, N. C., Mante, V., Simoncelli, E. P. and Movshon, J. A. (2006). How MT cells analyze the motion of visual patterns. Nature Neuroscience **9**: 1421-1431.

Simoncelli, E. P. and Heeger, D. J. (1998). A model of neuronal responses in visual area MT. Vision Research **38**: 743-761.

Smith, M. A., Majaj, N. J. and Movshon, J. A. (2005). Dynamics of motion signaling by neurons in macaque area MT. Nature Neuroscience **8**: 220-228.

Srinivasan, M., Lehrer, M., Kirchner, W. and Zhang, S. (1991). Range perception through apparent image speed in freely flying honeybees. Visual Neuroscience **6**: 519-535.

Tsui, J. M., Hunter, J. N., Born, R. T. and Pack, C. C. (2010). The role of V1 surround suppression in MT motion integration. Journal of Neurophysiology **103**: 3123-3138.

Van Essen, D. C. and Maunsell, J. H. (1983). Hierarchical organization and functional streams in the visual cortex. Trends in Neurosciences **6**: 370-375.

Van Santen, J. P. and Sperling, G. (1985). Elaborated reichardt detectors. Journal of the Optical Society of America A **2**: 300-321.

Von der Heydt, R., Peterhans, E. and Baumgartner, G. (1984). Illusory contours and cortical neuron responses. Science **224**: 1260-1262.

Zarei Eskikand, P., Kameneva, T., Burkitt, A. N., Grayden, D. B. and Ibbotson, M. R. (2019). Pattern motion processing by MT neurons. Frontiers in Neural Circuits **13**: 43-59.

Zarei Eskikand, P., Kameneva, T., Burkitt, A. N., Grayden, D. B. and Ibbotson M. R. (2020). Adaptive surround modulation of MT neurons: A computational model. Frontiers in Neural Circuits **14**: 529345.

Zarei Eskikand, P., Kameneva, T., Ibbotson, M. R., Burkitt, A. N. and Grayden, D. B. (2016). A possible role for end-stopped V1 neurons in the perception of motion: a computational model. PloS One **11**: e0164813.

Zarei Eskikand, P., Kameneva, T., Ibbotson, M. R., Burkitt, A. N. and Grayden, D. B. (2018). A biologically-based computational model of visual cortex that overcomes the X-junction illusion. Neural Networks **102**: 10-20.

Zeki, S. (1971). Convergent input from the striate cortex (area 17) to the cortex of the superior temporal sulcus in the rhesus monkey. Brain Research **28**: 338-340.

Zeki, S. M. (1974). Functional organization of a visual area in the posterior bank of the superior temporal sulcus of the rhesus monkey. The Journal of Physiology **236**: 549-573.